\documentclass[12pt]{article}
\usepackage{amsmath}
\usepackage{amscd}
\usepackage{amssymb}
\usepackage{array}

\begin{document}
\rightline{CERN-TH/2003-037}

\newcommand{\id}{\relax{\rm 1\kern-.28em 1}}

\newcommand{\R}{\mathbb{R}}
\newcommand{\C}{\mathbb{C}}
\newcommand{\Z}{\mathbb{Z}}

\newcommand{\K}{\mathbf{K3}}
\newcommand{\T}{\mathbf{T}}

\newcommand{\rSO}{\mathrm{SO}}
\newcommand{\rGL}{\mathrm{GL}}
\newcommand{\rSL}{\mathrm{SL}}
\newcommand{\rSU}{\mathrm{SU}}
\newcommand{\rSp}{\mathrm{Sp}}
\newcommand{\rUSp}{\mathrm{USp}}
\newcommand{\rU}{\mathrm{U}}

\newcommand{\fsu}{\mathfrak{su}}
\newcommand{\fso}{\mathfrak{so}}
\newcommand{\fh}{\mathfrak{h}}
\newcommand{\fp}{\mathfrak{p}}

\newcommand{\cM}{\mathcal{M}}
\newcommand{\cN}{\mathcal{N}}
\newcommand{\cL}{\mathcal{L}}
\newcommand{\cH}{\mathcal{H}}

\vskip 1cm

  \centerline{\LARGE \bf 4-D gauged supergravity analysis
  }

  \bigskip

  \centerline{\LARGE \bf  of Type IIB vacua on  $\K\times \T^2/\Z_2$
  }

 \vskip 1.5cm
\centerline{L. Andrianopoli$^\flat$, R. D'Auria$^\sharp$,  S.
Ferrara$^{\flat \star}$ and M. A. Lled\'o$^\sharp$.}

\vskip 1cm

\centerline{\it $^\flat$ CERN, Theory Division, CH 1211 Geneva 23,
Switzerland.}

\bigskip

\centerline{\it $^\sharp$ Dipartimento di Fisica, Politecnico di
Torino,} \centerline{\it Corso Duca degli Abruzzi 24, I-10129
Torino, Italy  and  } \centerline{\it   INFN, Sezione di Torino,
Italy. }

\bigskip

\centerline{$^\star$\it INFN, Laboratori Nazionali di Frascati,
Italy}

\vskip 1.5cm

\begin{abstract}
We analyze $N=2,1,0$ vacua of type IIB string theory on $\K\times
\T^2/\Z_2$ in presence of three-form fluxes from a four
dimensional supergravity viewpoint. The quaternionic geometry of
the $\K$ moduli space together with the special geometry of the NS
and R-R dilatons and of the $\T^2$-complex structure moduli play a
crucial role in the analysis. The introduction of fluxes
corresponds to a particular gauging of $N=2$, $D=4$ supergravity.
Our results agree with a recent work of Tripathy and Trivedi. The
present formulation shows the power of supergravity  in the study
of effective theories with broken supersymmetry.
\end{abstract}

\vfill\eject

\section{Introduction}

Recently, compactifications of string and M-theories in presence
of $p$-form fluxes have received much attention. They give rise to
effective models where the moduli stabilization is a simple
consequence of a combined Higgs and super Higgs mechanism
\cite{ps}-\cite{tt}.

Interestingly enough, many of these models are examples of
no-scale supergravities  \cite{cfkn}--\cite{ckpdfwg} in the
context of string and M-theory \cite{tv}\cite{cklt}--\cite{lm}.
Their low energy effective limits can be understood in terms  of
gauged supergravities, which, for a special choice of the gauging,
allow partial supersymmetry breaking without a cosmological term
\cite{adflf,adfld}\cite{cgp}--\cite{tz}.

A particularly appealing class of examples is obtained by
considering type IIB theory compactified on orientifolds such as
$\T^6/\Z_2$ \cite{fp,kst,dfv,dflv,adflq} and $\K\times
\T^2/\Z_2$\cite{tt}, or Calabi-Yau manifolds
\cite{ps,tv,ma,lm,da}, in presence of three form fluxes. When
compactifiying on the orientifold $\T^6/\Z_2$ with fluxes, one
finds vacua with reduced supersymmetry $N=3,2,1,0$ \cite{fp,kst,
dfv,dflv}, while in the case of $\K\times \T^2/\Z_2$, one obtains
vacua with $N=2,1,0$ supersymmetry \cite{tt}.

 It is the aim of the present investigation to obtain a gauged supergravity interpretation
 of the $N=2,1,0$ vacua recently found by Tripathy and Trivedi
 \cite{tt} for the $\K\times \T^2/\Z_2$ theory.
In absence of fluxes we obtain an ungauged $N=2$ supergravity with
a certain content of hypermultiplets and vector multiplets
\cite{wlp,adcdffm,dfv,dflv}. Moreover, the underlying special and
quaternionic geometries for these multiplets \cite{wlp,baw} is
determined by the properties of the moduli spaces.  The
introduction of fluxes is then equivalent to gauge some isometries
of the quaternionic manifold by some of the vectors at our
disposal in the theory.

$N=2$ and $N=1$ vacua stabilize many of the moduli and correspond
to two different gaugings: they differ in the choice of
quaternionic isometries and in the choice of vectors which realize
the gauging.

Let $f$ denote generically the fermions of the theory, and let
$\epsilon$ be a rigid supersymmetry  (constant) parameter in four
dimensional Minkowski space. A Poincar\'e invariant configuration
must have all fields equal to zero except for the scalar fields,
which can be set to constants. This configuration has an unbroken
supersymmetry $\epsilon$ if the values of the scalar fields are
such that
\begin{equation}\delta_\epsilon f=0.\label{vacuumeq}\end{equation}  The
crucial fact is that these are necessary and sufficient conditions
for the configuration to be a supersymmetric vacuum (with unbroken
supersymmetry $\epsilon$)  with vanishing vacuum energy.

It is our purpose to find solutions to (\ref{vacuumeq}) in the low
energy effective $N=2$ supergravity derived from compactifications
of the type IIB theory on $\K\times \T^2/\Z_2$ in presence of
fluxes. A similar analysis for the $\T^6/\Z_2$ theory has been
done in Refs. \cite{dfv,dflv,adflq}.

\section{Type IIB superstring on a $\K\times \T^2/\Z_2$ orientifold}

Type IIB compactified on $\K\times \T^2$ has been widely studied
in the context of the  type IIA-type IIB and type I-heterotic
string dualities \cite{ya,fhsv,klm,dlr,as}.

 The bulk sector of this theory is largely based on
properties of the moduli space  \cite{se,cfg} of the $\K$ manifold
\cite{as} and the torus $\T^2$. Before the orientifold projection
this theory has $N=4$ supersymmetry. After the $\Z_2$ projection,
and in absence of fluxes, the theory has $N=2$ supersymmetry. We
discuss the spectrum of the projected theory \cite{tt}. It
consists of the following multiplets:
\smallskip

\noindent 1. the graviton multiplet, $\quad[ (2), 2(\frac 3 2),
(1)]$,

\smallskip

\noindent 2. three vector multiplets, $\quad 3\times [ (1),
2(\frac 1 2), 2(0)]$,

\smallskip

\noindent 3. twenty hypermultiplets, $\quad 20\times [2(\frac 1
2), 4(0)]$.

\bigskip

We count first the scalar degrees of freedom remaining after the
projection. The internal manifold is parametrized by a pair of
complex coordinates on the $\K$ factor (indexed by $L=1,2$) and a
pair of real coordinates on the torus (indexed by $i=1,2$).  The
metric of the internal manifold is the direct product metric.  The
moduli space of the metrics on $\K$ is, up to a quotient by
discrete transformations,
\begin{equation}\frac{\rSO(3,19)}{\rSO(3)\times
\rSO(19)}\times \R^+_\K,\label{k3metrics}\end{equation} and has
dimension 58. We denote by $\rho_2$ the parameter corresponding to
$\R^+_\K$.

 The moduli space of the metrics on $\T^2$ is
$$\frac{\rSL(2,\R)}{\rSO(2)}\times \R^+_{\T^2},$$
of dimension 3 and parametrized by the Kaehler modulus $\phi$ and
the complex structure $\tau$:
\begin{eqnarray}
&\phi=\sqrt{g}=\sqrt{g_{11}g_{22}-{g_{12}}^2}\qquad &\hbox{1 real scalar}\label{kmodulus}\\
&\tau=\tau_1+ i\tau_2, \quad \tau_1=\frac {g_{12}} {g_{22}}, \;
\tau_2=\frac {\sqrt{g}} {g_{22}},\qquad &\hbox{1 complex
scalar}.\end{eqnarray}

Let $C_{\mu\nu\rho\sigma}$ be the four form field of type IIB,
with self-dual field strength. There are 23 RR real scalars that
come from this form when the indices are taken along the internal
manifold. The massless modes correspond to cohomology classes. The
Hodge numbers of odd order on $\K$ are zero. For  even order, the
only non vanishing ones are  $h_{0,0}=h_{2,2}=1$,
 $h_{2,0}=h_{0,2}=1$ and $h_{1,1}=20$. The
torus $\T^2$ has Betti numbers $b_2=b_0=1$ and $b_1=2$.  Then, the
components of $C_{\mu\nu\rho\sigma}$ that will give rise to scalar
fields are of the following forms:
\begin{eqnarray}&
C_{LMij}=C\epsilon_{LM}\epsilon_{ij},\;  &\hbox{one complex scalar}\label{scalar1}\\
& C_{L\bar M ij}=C_{L\bar
M}\epsilon_{ij},\;   &\hbox{20 real scalars}\label{scalar2}\\
& C_{L\bar P M\bar Q}= \rho_1\epsilon_{L M}\epsilon_{\bar P\bar
Q},\;   &\hbox{1 real scalar.}\label{scalar3}\end{eqnarray}

Finally, we have the two type IIB dilatons that  give two scalars
in four dimensions. We denote them by $\varphi_0$ and $C_0$.

\bigskip

The manifold of the $\K$ metrics (\ref{k3metrics}) with $\R_\K^+$
replaced by  $\R_\T^2$ (see below for an explanation), enlarges
with the 22 scalars of (\ref{scalar1}) and (\ref{scalar2}) to the
quaternionic manifold \cite{se}
$$\frac{\rSO(4,20)}{\rSO(4)\times \rSO(20)}.$$
This manifold has real dimension 80, corresponding to the 80
scalars of the twenty hypermultiplets.

To understand the assignment of the $\R^+$ factors to the scalar
manifolds, let us look at the kinetic term for the two-form field
strengths as they come from ten dimensions (up to a factor
depending on the dilaton),
$$\sqrt{g_{10}}g^{\hat\mu_1\hat\nu_1}g^{\hat\mu_2\hat\nu_2}g^{\hat\mu_3\hat\nu_3}H_{\hat\mu_1\hat\mu_2\hat\mu_3}
H_{\hat\nu_1\hat\nu_2\hat\nu_3}.$$ When compactifying on $\K\times
\T^2$, the volume factorizes as
$\sqrt{g_{10}}=\sqrt{g_{4}}\sqrt{g_{2}}\nu$, where $\nu$ is the
volume on $\K$ and $\sqrt{g_2}$ the volume on $\T^2$. Since the
bulk vectors in $D=4$ arise by taking an index along $\T^2$, the
relevant term is
$$\sqrt{g_{4}}\sqrt{g_{2}}\nu g^{\mu_1\nu_1}g^{\mu_2\nu_2}g^{ij}H_{\mu_1\mu_2i}
H_{\nu_1\nu_2j}.$$ The factor $\sqrt{g_{2}}g^{ij}$ is conformally
invariant in dimension 2 and depends only on the complex structure
of the torus $\tau$, while the modulus $\nu=\rho_2$ appears
explicitly.  Also, if one includes D7 brane gauge fields, their
four dimensional coupling depends on the $\K$ volume but not on
the $\T^2$ Kaehler modulus \footnote{We acknowledge an
enlightening conversation with C. Angelantonj on this point.}.
Then the coordinate $\rho_2$ of the $\K$ volume seats in a vector
multiplet.  These couplings are insensitive to the rescaling of
the Einstein-Hilbert term.

The three complex scalars of the  vector multiplets are
\begin{eqnarray} &\rho=\rho_1-i\rho_2,\quad &\Im\rho < 0\nonumber\\
&\tau=\tau_1 +i\tau_2, \quad &\Im\tau >0\nonumber\\
&\sigma=C_0+ie^{\varphi_0}, \quad &\Im
\sigma>0\label{imaginario}\end{eqnarray} parametrizing the coset
$$\frac {\rSL(2,\R)}{\rSO(2)}\times \frac
{\rSL(2,\R)}{\rSO(2)}\times \frac {\rSL(2,\R)}{\rSO(2)}.$$

The fact that the $\T^2$ complex structure moduli are in a vector
multiplet can also be understood by considering that type IIB on
the orientifold $\K\times \T^2/\Z_2$ is a truncation to $N=2$ of
$N=4$ supergravity corresponding to the compactification of type
IIB on $\K\times \T^2$, whose moduli space is
$$\frac {\rSU(1,1)}{\rU(1)}\times \frac
{\rSO(6,22)}{\rSO(6)\times \rSO(22)}.$$ There, the complex
structure moduli are in the factor ${\rSU(1,1)}/{\rU(1)}$, which
is a Kaehler-Hodge manifold \cite{dlr}.

\subsection{The quaternionic manifold ${\rSO(4,20)}/{\rSO(4)\times
\rSO(20)}$ as a fibration over ${\rSO(3,19)}/{\rSO(3)\times
\rSO(19)}$}

The coset $\cM_q={\rSO(4,20)}/{\rSO(4)\times \rSO(20)}$ is the
symmetric space associated to the Cartan decomposition
$$\fso(4,20)=\fh+ \fp, \qquad \mathrm{with}\quad \fh=\fso(4)+\fso(20)\quad \mathrm{and} \quad \fp=
(\mathbf{4},\mathbf{20}).$$ We want to give a local
parametrization of this coset which displays
$\cN={\rSO(3,19)}/({\rSO(3)\times \rSO(19)})$ as a submanifold of
$\cM_q$. It can indeed be proven that globally, $\cM_q$ is a
fibration over $\cN$ \cite{st}.

Consider the following decomposition of the Lie algebra
$\fso(4,20)$:
\begin{equation}\fso(4,20)=\fso(3,19)+\fso(1,1)+
(\mathbf{3},\mathbf{19})^++(\mathbf{3},\mathbf{19})^-.\label{decomp}\end{equation}
According to this decomposition, we can find a local
parametrization of  $\cM_q$ via the following element (coset
representative) of ${\rSO(4,20)}$:
\begin{equation}G=\mathrm{e}^{C^IZ_I}\mathrm{e}^{\phi S}{L},\label{parametrization}\end{equation} where
$\{Z_I\}_{I=1}^{22}$ is a set of generators of the abelian
subalgebra $(\mathbf{3},\mathbf{19})^+$ in (\ref{decomp}), $S$ is
the generator of $\fso(1,1)$ and $L$ is a coset representative of
$\cN$, $L\in \rSO(3,19)$. It is given in terms of 57 parameters
$e_m^a$, $m=1,2,3$, $a=1,\dots 19$ as
$$L=\begin{pmatrix}(1+ee^T)^\frac 1 2&e\\e^T& (1+e^Te)^{\frac 12}\end{pmatrix}.$$
Because of the action of $\rSO(4,20)$, there is a submanifold
parametrized by the coordinates $C^I$ with the topology of
$S^3\times S^{19}/\Z_2$.

In this parametrization, the Maurer-Cartan form is simply
\begin{equation}G^{-1}dG=\mathrm{e}^\phi (L^{-1})^I_JdC^JZ_I+ d\phi S
+L^{-1}dL.\label{vielbein}\end{equation} The connection and the
vielbein 1-forms are the projections of the Maurer-Cartan form
over the spaces $\fh$ and $\fp$ respectively. In fact, since  the
fundamental representation of $\rSO(n,m)$ is real, these
projections correspond to the symmetric and antisymmetric parts of
the matrices,
\begin{eqnarray*}&(G^{-1}dG)_{\mathrm{antisym}}=(G^{-1}dG)_{\fh}
&\hbox{is the connection 1-form},\\
&(G^{-1}dG)_{\mathrm{sym}}=(G^{-1}dG)_{\fp} &\hbox{is the vielbein
1-form}. \end{eqnarray*}

We want to write explicitly the vielbein one-form. We take $G$ in
the fundamental representation of $\rSO(4,20)$,
$G(q)^\Lambda_\Sigma$, with $q^u$ the coordinates on the
quaternionic manifold. Since $G(q)^\Lambda_\Sigma$ is a coset
representative, we will be interested in the transformation
properties to the right with respect to the subgroup
$H=\rSO(4)\times \rSO(20)$, and will denote it as
$G(q)^\Lambda_\sigma$. The Maurer-Cartan form will have indices in
the same representation of $\fh$,
$$(G^{-1}dG)^\lambda_\sigma, \qquad \lambda,\sigma=1,\dots 24.$$
Notice that $H=\rSU(2)_R\times
\rSU(2)\times\rSO(20)\subset\rSU(2)_R\times\rUSp(40)$, where
$\rSU(2)_R$ is the R-symmetry group and $\rSU(2)_R\times\rUSp(40)$
is the holonomy group of a general quaternionic manifold.

Since we are interested in showing explicitly the symmetry under
$\rSO(3,19)$, we decompose
$$\begin{CD}(\mathbf{4+20})@>>\rSO(3)\times \rSO(19)>(\mathbf{3+19})+\mathbf{1}+\mathbf{1}\end{CD}.$$
In this decomposition, the diagonal group  $(\rSU(2)_R\times
\rSU(2))_\mathrm{diag}$ in $H$ is identified with the first factor
$\rSO(3)$.

For the vielbein we have,
\begin{equation}
(G^{-1}dG)_{\mathrm{sym}} =U_udq^u=\begin{pmatrix} 0&P^m_b&U^m&0\\
(P^t)^a_n&0&0&V^a\\
({U^t})_n&0&0&d\phi\\
0&{V^t}_b&d\phi&0\end{pmatrix},\label{explvielbein}
\end{equation}
with $P^m_b = \left(L^{-1} dL\right) ^m_a$ the vielbein of the
scalar manifold ${\rSO(3,19)}/({\rSO(3) \times \rSO(19)})$ and
\begin{eqnarray}
 U^m=\mathrm{e}^\phi\left[ [(\id + e\cdot e^t)^{\frac 12}]^m_n
dC^n
+ e^m_a dC^a \right] \label{um}\\
V^a=\mathrm{e}^\phi\left[ e^a_m dC^m + [(\id + e^t\cdot e)^{\frac
12}]^a_b dC^b \right]
\end{eqnarray}
Notice that the space $\fp$ corresponds to the $(\mathbf{4,20})$
representation of $H$, which decomposes as
$$\begin{CD}(\mathbf{4,20})@>>\rSO(3)\times \rSO(19)>(\mathbf{3,19})+(\mathbf{3,1})+(\mathbf{1, 19})+ (\mathbf{1, 1})\end{CD}.$$
and each block in (\ref{explvielbein}) corresponds to one of
these.

\bigskip

We note  that the components on the generators $Z_I$ of the
Maurer-Cartan form $$\mathrm{e}^\phi (L^{-1})^I_JdC^J$$ contribute
both to the vielbein and to the connection of the quaternionic
manifold $\cM$. In particular, the contribution of this term to
the $\rSU(2)_R$-connection is proportional to
\begin{equation}\omega_I^xdC^I\propto \mathrm{e}^\phi
(L^{-1})_I^xdC^I=\mathrm{e}^\phi(1+ee^T)_m^xdC^m-\mathrm{e}^\phi
e_a^xdC^a, \qquad x=1,2,3. \label{omega} \end{equation}
 These
formulae will be useful in the calculation of the scalar
potential.

\subsection{\label{vectormultiplets}Vector multiplets and special geometry}
The vector multiplet moduli space is
$$\cM_{n_v=3}=\frac {\rSL(2,\R)}{\rSO(2)}\times \frac
{\rSL(2,\R)}{\rSO(2)}\times \frac {\rSL(2,\R)}{\rSO(2)}.$$ It is a
special Kaehler-Hodge manifold of the series \cite{ckpdfwg,cp}
$$\cM_{n_v}=\frac {\rSL(2,\R)}{\rSO(2)}\times \frac
{\rSO(2,n_v-1)}{\rSO(2)\times \rSO(n_v-1)}$$ with $n_v=3$.

Let $\cL$ be the Hodge line bundle. The special geometry
\cite{str,cdf,cedf} of $\cM_{n_v}$ consists on a holomorphic
vector bundle $\cH$ with structure group $\rSp(2+2n_v,\R)$ and a
global section
$$\Omega=\begin{pmatrix}
X^\Lambda(z)\\F_\Lambda(z)\end{pmatrix}$$ on $\cH\otimes \cL$ such
that the Kaehler form $J$ is given by
$$J=-\frac i {2\pi} \partial\bar \partial\ln i[\bar X^\Lambda F_\Lambda-\bar F_\Lambda X^\Lambda].$$
In an open set the Kaehler potential is given by
$$K=-\ln i[\bar X^\Lambda F_\Lambda-\bar F_\Lambda X^\Lambda].$$
 In a point $z$ of the intersection of two open sets the section
transforms as
$$\begin{pmatrix}
X^\Lambda(z)\\F_\Lambda(z)\end{pmatrix}=
\mathrm{e}^{f(z,A,B,C,D)}\begin{pmatrix}
A^\Lambda_{\Lambda'}&B^{\Lambda\Lambda'}\\C_{\Lambda\Lambda'}& D
_\Lambda^{\Lambda'}\end{pmatrix}\begin{pmatrix}
X^{\Lambda'}(z)\\F_{\Lambda'}(z)\end{pmatrix},$$ where  $({A,B,C,
D})$ define a constant, symplectic transformation:
$$A^TD-C^TB=\id, \quad A^TC=C^TA, \quad B^TD=D^TB,$$ and
$f(z,A,B,C,D)$ is a holomorphic phase of the Hodge bundle.

\bigskip

From the doublet of two forms in IIB, $B_{\mu\nu}^\alpha$, one can
obtain vector fields in four dimensions when one of the indices is
taken over the torus $\T^2$ (the odd cohomology of $\K$ is zero).
Then the four gauge fields  in four dimensions are \cite{tt}
$$A_\mu^{\alpha i}= B_{\mu i}^\alpha.$$
Therefore they are in the  representation  $(\frac 1 2, \frac 1
2)$ of the type IIB R-symmetry $\rSO(2,1)\simeq\rSL(2, \R)$, times
the $\rSL(2,\R)$ associated to the $\T^2$ complex structure. In
the homomorphism $\rSO(2,1)\times\rSO(2,1)\simeq\rSO(2,2)$ the
$(\frac 1 2, \frac 1 2)$ goes to the fundamental representation,
so we can set an index $\Lambda=(i,\alpha)=0,1,2,3$ and denote the
gauge fields as $A^\Lambda_\mu$.

We have therefore to choose a symplectic embedding of $\rSO(2,2)
\times \rSL(2,R)$ in $\rSp(8, R)$ such that $\rSO(2,2)$ is an
electric subgroup. The third $\rSL(2,R)$ instead acts on the
vectors as an electric-magnetic duality.

Let $\eta=\mathrm{diag}(+1,+1,-1,-1)$. $\rSO(2,2)$ is embedded in
$\rSp(8, R)$ as
$$\begin{pmatrix}
A&B\\C& D\end{pmatrix},\qquad A\in \rSO(2,2), \quad D=(A^T)^{-1},
\quad B=C=0,$$ while the  third  $\rSL(2,\R)$ is embedded as
$$\begin{pmatrix}
a\id&b\eta\\c\eta& d\id\end{pmatrix},\qquad ad-bc=1.$$

In order to have the symplectic embedding chosen above explicitly
manifest in the theory, one has to choose a local frame for the
symplectic bundle. In this frame, the global section $\Omega$ of
the special geometry is given by \cite{cdfp}:
$$\Omega=\bigl(X^\Lambda(\sigma,\tau),
F_\Lambda=\rho\eta_{\Lambda\Sigma}X^\Sigma(\sigma,\tau)\bigr),$$
with $$X^\Lambda X^\Sigma\eta_{\Lambda\Sigma}=0, \qquad X^\Lambda
\bar X^\Sigma\eta_{\Lambda\Sigma}=\mathrm{e}^{-\hat K},$$ and
where $\hat K= \frac 1 2 i(\bar \tau-\tau)i(\bar \sigma-\sigma)$
is the Kaehler potential of the submanifold
$\rSO(2,2)/(\rSO(2)\times \rSO(2))$. The explicit dependence of
$X^\Lambda$ in terms of the local coordinates $(\rho,\sigma,\tau)$
is
\begin{equation}X^0=\frac 1 2 (1-\sigma\tau), \quad X^1=-\frac 1 2
(\sigma+\tau),\quad X^2=-\frac 1 2 (1+\sigma\tau), \quad X^3=\frac
1 2 (\tau-\sigma).\label{ixlambda}\end{equation}

It is important to notice that in this embedding $F_\Lambda$
cannot be written as $\partial_\Lambda F(X)$. The prepotential
$F(X)$ does not exist. This allows to have partial breaking of
$N=2$ supersymmetry \cite{fgp}, otherwise impossible \cite{cgp}.

The Kaehler potential of $\cM_{n_v=3}$ is given by the formula
\begin{equation}\mathrm{e}^{-K}=i(\bar X^\Lambda F_\Lambda-X^\Lambda\bar
F_\Lambda)=\frac i 2 (\rho-\bar\rho)i(\bar\tau
-\tau)i(\bar\sigma-\sigma)=\mathrm{e}^{-\tilde K}
\mathrm{e}^{-\hat K},\label{kaehlerpot}\end{equation} where
$\mathrm{e}^{-\tilde K}= i(\rho-\bar\rho).$

From the symplectic section we get the following kinetic matrix
for the vectors \cite{cdfp,adcdffm}
$$\mathcal{N}_{\Lambda\Sigma}=(\rho-\bar \rho)(\phi_\Lambda\bar
\phi_\Sigma+\bar\phi_\Lambda \phi_\Sigma) +\bar
\rho\eta_{\Lambda\Sigma}, \qquad \phi^\Lambda=\frac
{X^\Lambda}{(X^\Sigma\bar X_\Sigma)^{\frac 1 2}},$$ which has the
important property
$$-\frac 1 2 (\Im\mathcal{N}_{\Lambda\Sigma})^{-1}-4\bar
L^{(\Lambda}L^{\Sigma)}=-\frac {\eta^{\Lambda\Sigma}}{i(\rho-\bar
\rho)}.$$

\section{\label{gaugingtraslational}Gauging of the traslational isometries and fluxes}

The gauging of the present theory  involves the four abelian
vectors and the scalars of the quaternionic manifold, in
particular the axions of the 22 abelian isometries, whose Killing
vectors correspond in the Lie algebra (\ref{decomp}) to a
Lorentzian vector of $\rSO(3,19)$. We can in principle gauge a
four dimensional subalgebra of this abelian algebra. As in
(\ref{parametrization}) we denote by $(C_m, C_a)$ with $m=1,2,3$
and $a=1,\dots 19$, the 22 axions. Their covariant derivatives are
\begin{eqnarray*} &D_\mu C_m=\partial_\mu
C_m+f_{m,\Lambda}A^{\Lambda}_\mu,\\&D_\mu C_a=\partial_\mu
C_a+h_{a,\Lambda}A^{\Lambda}_\mu,\end{eqnarray*} with
$\Lambda=0,\dots 3$.  $f_{m,\Lambda}$, $h_{a,\Lambda}$ are the
coupling constants. When performing the dimensional reduction, the
kinetic terms for the axions appear with these covariant
derivatives, and the coupling constants are related to the
three-form fluxes on $\K\times \T^2$.

More precisely, let us undo the relabeling of the $\rSO(2,2)$
vector index, $\Lambda =(i,\alpha)$, $i=1,2$ and $\alpha =1,2$ as
in Section \ref{vectormultiplets}, and the 22 vector indices
$(m,a)= (LM,L\bar M)$ as in (\ref{scalar1}),(\ref{scalar2}). Then
the coupling constants become
$$(\tilde f_{i,LM,\alpha},\tilde h_{i,L\bar M,\alpha}).$$
They can be identified with the three form fluxes with one index
on $\T^2$ and the other two on $\K$.

There are different choices for the coupling constants, according
to the supersymmetries that we want to have for the vacua:

\smallskip

\noindent $\bullet$ For configurations with $N=2$ supersymmetry we
will take $f_{m,\Lambda}=0$,  $h_{a,0}=h_{a,1}=0$ and
$h_{1,2}=g_2, h_{2,3}=g_3,$. The vectors that are ``higgsed" (that
acquire mass) are the vector partners of the IIB dilaton and the
$\T^2$ complex structure moduli.

\smallskip

\noindent $\bullet$ For configurations with $N=1,0$ supersymmetry
we will take, in the first place, $h_{a,\Lambda}=0$,
$f_{m,2}=f_{m,3}=0$ and only $f_{1,0}=g_0, f_{2,1}=g_1$ different
from zero.  In this case, the vectors that acquire mass are the
graviphoton and the vector partner of the $\K$ volume modulus.

\bigskip

\noindent This means that  the coupling constant
$\rSO(2,2)$-vector $g_\Lambda$ (with metric\\
$\mathrm{diag}(+,+,-,-)$) has negative norm for $N=2$
configurations and positive norm  for $N=0,1$ configurations. The
$N=1$ is obtained by imposing the further constraint,
$|g_0|=|g_1|.$

\bigskip

\noindent $\bullet$ For $N=2$ supersymmetry preserving vacua, no
other choices are allowed, while for configurations with $N=1,0$
supersymmetry there exists a more general choice, with all
couplings $g_\Lambda$ non vanishing. All the vectors acquire mass.
These configurations will be discussed separately in Section
\ref{noscale}.

\section{Supersymmetric vacua}
As a consequence of the supersymmetric Ward identities \cite{df}
one can obtain any supersymmetric configuration as a solution of
the constraints
$$\delta_{\epsilon}\lambda=0,$$
where $\epsilon$ is the constant parameter of the global unbroken
supersymmetry and $\lambda$ are the spin $\frac 1 2$ fields. This
can be done without looking to the explicit form of the potential.
If we look for vacua with a Poincar\'e symmetry, one has the
further constraint
$$\delta_{\epsilon}\psi_{\mu A}=0,$$
with $\psi_{\mu A}$ the two gravitino fields.

In $N=2$ supergravity, we have hypermultiplets and vector
multiplets. The fermion fields denote always the chiral
projections. We denote by $\psi_{\mu A}$ the gravitinos, with
$A=1,2$ refering to the $\rSU(2)$ R-symmetry; by $\lambda^{iA}$
the gauginos with $i=1, \dots n_v$, which form a contravariant
vector on the special Kaehler manifold $\cM_{n_v}$; and the
hyperinos by $\zeta_\alpha$, with $\alpha=1, \dots 2n_h)$ ($n_h$
is the number of hypermultiplets). The index alpha is a vector
index
 of $\rUSp(2n_h)$, which together with $\rSU(2)$ (index A=1,2)
form the reduced holonomy of the quaternionic manifold.

 We will denote   by $k_\Lambda=k_\Lambda^u\partial_u$  the  Killing
vectors of the translational isometries of  the
 quaternionic manifold, which will be gauged (hence with an index $\Lambda=0,\dots 3$ as the vectors).
  Their  prepotential  is denoted by  $P_\Lambda^x$, with $x=1,2,3$ (it  is an $\rSU(2)$ triplet).
  If $\Omega_{uv}$ is the
curvature two form, and $\Omega^x_{uv}$ its $\fsu(2)$ components,
then
$$k_{u\Lambda}=-\frac{1}6\Omega^x_{uv}\nabla^vP_\Lambda^x.$$
(The index $x$ is contracted with the Euclidean metric).

\medskip
In addition, we denote by $g_{i\bar j}$ the Kaehler metric of the
special manifold $\cM_{n_v}$, $K$ the Kaehler potential. In terms
of the holomorphic section $\Omega$ we can define
$$V=\mathrm{e}^{K/2}\begin{pmatrix}X^\Lambda\\F_\Lambda\end{pmatrix}$$
which is not holomorphic but is {\it covariantly holomorphic}
$$\mathcal{D}_{\bar i}V=(\partial_{\bar i} -\frac 1
2\partial_{\bar i}K)V=0.$$

The  supersymmetry transformations of the fermionic fields for a
constant parameter $\epsilon$  are as follows \cite{abcdffm}:

\begin{eqnarray} \delta\psi_{A\mu}&=&-\frac 1 2 P_\Lambda^xX^\Lambda
\mathrm{e}^{\frac K
2}(\sigma^x)_{AB}\gamma_\mu\epsilon^B\label{susy1}\\
\delta\lambda^{iA}&=&ig^{i\bar j}\mathcal{D}_{\bar j}(\bar
X^\Lambda
\mathrm{e}^{\frac K 2})P_\Lambda^x(\sigma^x)^{AB}\epsilon_B\label{susy2}\\
\delta\zeta_{\tilde A a}&=& 2\varepsilon^{AB}U_{A\tilde A a,
u}k_\Lambda^u\bar
X^\Lambda\mathrm{e}^{\frac K 2}\epsilon_B\label{susy3}\\
\delta\zeta_{\tilde A }&=& 2\varepsilon^{AB}U_{A\tilde A ,
u}k_\Lambda^u\bar X^\Lambda\mathrm{e}^{\frac K
2}\epsilon_B\label{susy4}
\end{eqnarray}
where the hyperinos transformation laws are decomposed with
respect to the manifest holonomy $\rSO(3) \times \rSO(19)$, and
$$U_{A\tilde A a, u} = \varepsilon_{A\tilde A} V_{a,u}, \qquad
U_{A\tilde A, u} = (\sigma^m)_{A\tilde A} U_{m, u}$$
 with $U_m$, $V_a$ given in (\ref{explvielbein}).

\subsection{N=2 \label{n=2} supersymmetric configurations}

To have $N=2$ configurations with vanishing vacuum energy, the
variations of all fermions must vanish for any constant
supersymmetry parameter $\epsilon_A$ ($A=1,2$). This demands, from
equations (\ref{susy1}) and (\ref{susy2}), that
\begin{equation}
P^x_\Lambda =0 \qquad  \, x=1,2,3, \;\;\Lambda=0,1,2,3
\label{n2grav}
\end{equation}
 and, from the hyperinos variations (\ref{susy3}),
(\ref{susy4}), that
\begin{equation}
k^u_\Lambda X^\Lambda =0, \qquad \, u=1,\dots 80. \label{n2hyp}
\end{equation}
To preserve $N=2$ supersymmetry, the graviphoton $A^0_\mu$ cannot
acquire mass. We can switch on interactions for $A^\Lambda_\mu$,
with $\Lambda =2,3$, gauging two of the isometries associated to
the 19 axions $C^a$. We take the two Killing vectors $k_2^u$ and
$k_3^u$ whose only  non vanishing components are
\begin{eqnarray*}&&k^u_2=g_2\neq 0 \quad \hbox{ for}\, q^u=C^{a=1},
\quad \mathrm{and}\\&& k^u_3=g_3\neq 0 \quad \hbox{ for}\,
q^u=C^{a=2},\end{eqnarray*} for arbitrary constants $g_2$ and
$g_3$.

  Inserting in equation
(\ref{n2hyp}), this implies
\begin{equation}
X^2(\sigma,\tau ) = X^3(\sigma,\tau )=0\label{n2stabilize}
\end{equation}
Equation (\ref{n2stabilize}) stabilizes the two vector-multiplets
moduli $(\sigma,\tau)$, since from (\ref{ixlambda}) we have:
$$\tau =\sigma \, , \sigma^2 =-1 \,\quad \Longrightarrow \quad \sigma=\tau
=i.$$ Let us note that to have $N=2$ preserving vacua it is not
possible to gauge more than two vectors, since it would give extra
constraints on the $X^\Lambda$, incompatible with
(\ref{n2stabilize}) in the given symplectic frame
(\ref{ixlambda}). Also, the two vectors that realize the gauging
have to be $A^2_\mu$ and $A_\mu^3$, which are in the same
multiplets as the coordinates $\sigma$ and $\tau$. Indeed, it is
easy to see that this choice is the only one stabilizing the
moduli compatibly with the conditions $\Im(\sigma), \Im(\tau)>0$
of (\ref{imaginario}). This same result is obtained in Ref.
\cite{tt}, Section 5 with topological arguments.

 Equation (\ref{n2grav}) is solved by
recalling that, for gauged axion symmetries \cite{mi,da}, the
expression for the prepotential $P^x_\Lambda$ gets simplified to
$$P^x_\Lambda = \omega^x_u k^u_\Lambda ,$$
where $\omega^x_u $ is the $\rSU(2)_R$-connection. In this case we
have
$$P^x_\Lambda = \sum_{a=1}^2\omega^x_a k^a_\Lambda =
\sum_{a=1}^2\mathrm{e}^\phi e^x_a k^a_\Lambda\,, \quad\mbox{ for }
\Lambda=2,3$$ where we have used equation (\ref{omega}).

$P^x_\Lambda=0$ then implies
$$e^x_a =0\qquad \mbox{ for } a=1,2.$$

The $C^a$, $a=1,2$ are Goldstone bosons. They disappear form the
spectrum, making massive the gauge vectors $A_2$ and $A_3$. In
fact,  two of the original massless hypermultiplets (corresponding
to the degrees of freedom $C^a$ and $e_a^x$ for $a=1,2$) and the
two  vector multiplets of $A_2$ and $A_3$, combine into two long
massive vector multiplets $[1,4(\frac 1 2), 5(0)]$.

We see that the $N=2$ configurations are just an example of the
Higgs phenomenon of two vector multiplets. The residual moduli
space is
$$\frac{\rSO(4,18)}{\rSO(4)\times \rSO(18)}\times
\frac{\rSU(1,1)}{\rU(1)}.$$ The ${\rSU(1,1)}/{\rU(1)}$ factor
contains the $\K$ volume modulus, appertaining to  the remaining
massless vector multiplet. The moduli corresponding to the $\K$
metrics form the submanifold
$$\frac{\rSO(3,17)}{\rSO(3)\times \rSO(17)}\times \R,$$
in accordance with Ref. \cite{tt}.

\subsection{N=1 supersymmetric configurations}

In the $N=1$ supersymmetric vacua, the  graviphoton and the vector
partner of $\K$ volume modulus are gauged. Indeed, in  any
truncation of $N=2\rightarrow N=1$ supergravity with Poincar\'e
vacuum, the graviphoton must become massive \cite{adf}. The charge
vector $g^\Lambda$ (in  the notation of Section
\ref{gaugingtraslational}) can be chosen with components
$f_{1,0}=g^0\neq 0$, $f_{2,1}=g^1\neq 0$ and the rest zero. This
means that  we switch on the charges of the isometries associated
to two of the three axions $C^m$.

The relevant Killing vectors are \begin{eqnarray*} k_0^u=g_0\neq 0
\quad\mathrm{for}\;\; q^u=C^{m=1}\\
k_1^u=g_1\neq 0 \quad\mathrm{for}\;\; q^u=C^{m=2}
\end{eqnarray*}
The quaternionic prepotential for constant  Killing vectors is
$$P^x_{\Lambda}=\omega_u^xk^u_\Lambda, \qquad \Lambda=0,1,$$
so
$$P^x_0=\omega^x_{C_1}g_0, \qquad P^x_1=\omega^x_{C_2}g_1.$$
Using equation (\ref{omega}), we have
$$P_0^x=\mathrm{e}^\phi{(1+ee^t)^{\frac 1 2}}_{1}^xg_0, \qquad
P_1^x=\mathrm{e}^\phi{ (1+ee^t)^{\frac 1 2}}^x_{2}g_1.$$

We want to study vacua that preserve one supersymmetry (we choose
$\epsilon_2$). We have to impose that $\delta_{\epsilon_2}f=0$ in
equations (\ref{susy1}-\ref{susy4}).

From the variation of the antichiral hyperinos we have
(\ref{susy3},\ref{susy4})
\begin{eqnarray*}\delta_{\epsilon_2}\zeta^{\tilde A a}&=&0\;
\Longrightarrow
V^a_{\,m}k^m_\Lambda X^\Lambda=0, \quad \Lambda=0,1,\; m=1,2,\\
\delta_{\epsilon_2}\zeta^{\tilde A }&=&0\; \Longrightarrow
{\sigma^m}^{\tilde A }_BU_{m\,n}k^n_\Lambda X^\Lambda=0, \quad
\Lambda=0,1,\; m=1,2,3,\, n=1,2.\end{eqnarray*} The first of these
equations implies \begin{equation}e_m^a=0, \;
m=1,2\label{hyperino},\end{equation} and the second one turns out
to be proportional to the equation for the variation of the
gravitino. We solve it below.

\bigskip

From the variation of the gravitino (\ref{susy1}) we obtain
\begin{eqnarray*} \delta_{\epsilon_2}\psi_{1\mu}&=&0\;
\Longrightarrow S_{12}\propto P_\Lambda^xX^\Lambda\sigma^x_{1\,2}=0,\\
 \delta_{\epsilon_2}\psi_{2\mu}&=&0\;
\Longrightarrow S_{22}\propto
P_\Lambda^xX^\Lambda\sigma^x_{2\,2}=0,\end{eqnarray*} in terms of
 the mass matrix of the gravitinos: \footnote{ The sigma
matrices with the two indices down are $(\sigma_3, i\id,
\sigma_1)$.}
$$S_{AB}= \frac i 2 P_\Lambda^xX^\Lambda\mathrm{e}^{\frac K
2}(\sigma^x)_{AB} =\frac i 2 \mathrm{e}^{\frac K
2}(P_0^x\sigma^x_{AB}X^0+P_1^x\sigma^x_{AB}X^1).$$ Using
(\ref{hyperino}), $S_{AB}$ becomes proportional to
\begin{equation}
\mathrm{e}^{\phi}\mathrm{e}^{\frac K 2}
\begin{pmatrix}g_0X^0+ig_1X^1&0\\
0&-g_0X^0+ig_1X^1\end{pmatrix},\label{massmatrix}\end{equation}
and an $N=1$ invariant vacuum requires \begin{equation}
S_{22}\propto -g_0X^0+ig_1X^1=0.\label{s22}\end{equation}

\bigskip

Finally, from the (antichiral) gauginos variation we find
$$\delta_{\epsilon_2}(\lambda)^{\bar i}_A=0\Longrightarrow
P_\Lambda^x\mathcal{D}_j(X^\Lambda\mathrm{e}^{\frac K
2})(\sigma^x)_{A2}=0.$$ (Notice that we have used the complex
conjugate of the chiral gaugino $\lambda^{iA}$). The relevant
matrix is
$$\mathrm{e}^{\phi}\begin{pmatrix}g_0\mathcal{D}_i(X^0\mathrm{e}^{\frac K
2})+ig_1\mathcal{D}_i(X^1\mathrm{e}^{\frac K
2})&0\\
0&-g_0\mathcal{D}_i(X^0\mathrm{e}^{\frac K
2})+ig_1\mathcal{D}_i(X^1\mathrm{e}^{\frac K 2})\end{pmatrix},$$
The second term in the covariant derivative
$\mathcal{D}_i=(\partial_i +\frac 1 2
\partial_i K)$ gives a contribution proportional to
(\ref{massmatrix}). The first term gives the conditions
\begin{eqnarray*}
&&-g_0\partial_\sigma X^0+ig_1\partial_\sigma X^1=0,\\
&&-g_0\partial_\tau X^0+ig_1\partial_\tau X^1=0,\end{eqnarray*}
which imply
$$\tau=\sigma=i\frac {g_1}{g_0}.$$
(Note that $\partial_\rho X^\Lambda=0$, so there are no further
constraints.)

Then equation (\ref{s22}) gives
$$-g_0(1-\sigma^2)-2ig_1\sigma=0,$$
and by using $\sigma=ig_1/g_2$ we get
$$g_0^2=g_1^2,$$
which implies $\tau=\sigma=i$ and $g_0=g_1$ (the other possibility
$g_0=-g_1$ would give $\sigma$ and $\tau$ outside their domain of
definition.) Note that at this point $X^2=X^3=0$ as in the $N=2$
case. Also, this point is the self dual point of
$\rSL(2,\R)/\rSO(2)$ (fixed point of the transformation
$\tau\rightarrow -1/\tau$).

\bigskip

We now summarize the massless spectrum of the $N=1$ reduced
theory. From the 58 scalars of $\rSO(3,19)/(\rSO(3)\times
\rSO(19))\times\R^+_{\T^2}$ there remain 20 scalars parametrizing
$\rSO(1,19)/ \rSO(19)\times\R^+_{\T^2}$. From the 22 axions there
remain 20. All together they complete the scalars of 20 chiral
multiplets. The spectrum includes two massless vector multiplets
corresponding to $A^2_\mu$ and $A_\mu^3$ and an extra chiral
multiplet whose scalar field is $\rho$ from the $N=2$ vector
multiplet sector.

Models with $N=0$ also exist and can be studied by writing the
full $N=2$ potential. They can be also obtained by further gauging
the $N=2$ theory obtained  in section \ref{n=2},  or by adding a
superpotential to the $N=1$ theory. A complete description of the
non supersymmetric phases will be done elsewhere. In this paper we
will study the $N=0$ vacua which have vector charge with $g_{0},
g_1\neq 0$ or with all $g_\Lambda\neq 0$.

\section{Non supersymmetric vacua}
The study of $N=0$ vacua requires the knowledge of the potential
of the scalar fields, which can be computed, for an abelian
gauging, with the formula \cite{abcdffm}
$$V=4h_{uv}k^u_\Lambda k^v_\Sigma L^\Lambda\bar L^\Sigma+
(U^{\Lambda\Sigma}-3\bar L^\Lambda
L^\Sigma)P_\Lambda^xP_\Sigma^x,$$ where
$$U^{\Lambda\Sigma}=-\frac 1 2
(\Im\mathcal{N}_{\Lambda\Sigma})^{-1}-\bar L^\Lambda L^\Sigma,
\qquad L^\Lambda=\mathrm{e}^{\frac K 2}X^\Lambda=
\mathrm{e}^{\frac{\tilde K} 2}\mathrm{e}^{\frac {\hat K}
2}X^\Lambda,$$ and
$$\tilde K=-\ln i(\rho-\bar\rho), \qquad \hat K=-\ln\frac 12i(\bar \tau-\tau)i(\bar \sigma-\sigma).$$

The three contributions are the square of the supersymmetry
variations of the hyperinos, gauginos and gravitinos respectively.
The first two terms are positive definite while the last
contribution is negative definite.

In the model at hand the last two terms become
$$(U^{\Lambda\Sigma}-3\bar L^\Lambda
L^\Sigma)P_\Lambda^xP_\Sigma^x=-\mathrm{e}^{\tilde K}
\eta^{\Lambda\Sigma}P_\Lambda^xP_\Sigma^x,$$ with $\tilde K=-\ln
i(\rho-\bar\rho)$. Then the scalar potential becomes
\begin{eqnarray*}&V=4h_{uv}k^u_\Lambda k^v_\Sigma L^\Lambda\bar
L^\Sigma-\mathrm{e}^{\tilde K}
\eta^{\Lambda\Sigma}P_\Lambda^xP_\Sigma^x \\
&=4h_{uv}k^u_\Lambda k^v_\Sigma L^\Lambda\bar
L^\Sigma+\mathrm{e}^{\tilde K}(P_2^2+P_3^2)-\mathrm{e}^{\tilde
K}(P_0^2+P_1^2)\\&= \mathrm{e}^{\tilde K}\bigl[4h_{uv}k^u_\Lambda
k^v_\Sigma \mathrm{e}^{\hat K}X^\Lambda\bar
X^\Sigma+P_2^2+P_3^2-P_0^2-P_1^2 \bigr],\end{eqnarray*} where we
have used equation (\ref{kaehlerpot}).  Note that $V\geq 0$ if
$P_0=P_1=0$. In this case, all vacua with zero vacuum energy have
unbroken $N=2$ supersymmetry.

The potential can be computed by recalling that in the case of
gauged axion isometries of the quaternionic manifold we have
$$P_\Lambda^x=\omega_u^xk^u_\Lambda.$$
We will consider here only the case when the $C^m$ become charged
under the $\Lambda=0,1$ symmetries.

Recalling  equations (\ref{um}) and (\ref{omega}), we see that the
quaternionic metric $h_{uv}$ along the $C^m$ axion directions is
$$h_{m n}=\mathrm{e}^{2\phi}
(L^{-1})_m^I(L^{-1})_n^I=\mathrm{e}^{2\phi}(\delta_{mn}+2e_m^ae^a_n),$$
with $ I=1,\dots 22,\; m,n=1,2,3,\; a=1,\dots 19,$ while
$$\omega_m^x\omega_n^x=2\mathrm{e}^{2\phi}
(L^{-1})_m^x(L^{-1})_n^x=2\mathrm{e}^{2\phi}(\delta_{mn}+e_m^ae^a_n),\qquad
x=1,2,3.
$$
Therefore, the scalar potential is
$$V=\mathrm{e}^{2\phi}\mathrm{e}^{\tilde
K}\bigl[4(\delta_{mn}+2e_m^ae^a_n)k^m_\Lambda k^n_\Sigma
\mathrm{e}^{\hat K}X^\Lambda\bar
X^\Sigma-2(\delta_{mn}+e_m^ae^a_n)k^m_\Lambda
k^n_\Sigma\eta^{\Lambda\Sigma}\bigr].$$ By taking $k^1_0=g_0$ and
$k_1^2=g_1$ we have
\begin{eqnarray}V=&&\mathrm{e}^{2\phi}\mathrm{e}^{\tilde
K}\bigl[4\mathrm{e}^{\hat K}\bigl(g_0^2X^0\bar X^0+g_1^2X^1\bar
X^1+g_1 g_2(X^0\bar X^1+\bar X^0 X^1)
\nonumber\\&&+2e_1^ae_1^ag_0^2X^0\bar X^0+2e_2^ae_2^ag_1^2X^1\bar
X^1+2e_1^ae_2^ag_1 g_2(X^0\bar X^1+\bar X^0
X^1)\bigr)\nonumber\\&&-2(g_0^2+g_1^2+g_0^2e_1^ae^a_1+g_2^2e_2^ae^a_2)\bigr].\label{pot}\end{eqnarray}
We want now to compute the extrema of the potential. The
conditions
$$\frac {\partial V} {\partial\rho}=0, \qquad \frac {\partial V} {\partial\phi}=0$$
are satisfied at the points where $V=0$. We will see that this is
implied by the other extremum conditions. The equations
$$\frac {\partial V} {\partial\sigma}=0, \qquad \frac {\partial V} {\partial\tau}=0$$
are solved by
$$\mathcal{D}_\sigma X^0=\mathcal{D}_\tau X^0=\mathcal{D}_\sigma
X^1=\mathcal{D}_\tau X^1=0,$$ which gives
$$\sigma=\tau=i, \qquad \mathrm{e}^{\hat K}=\frac 1 2, \qquad  X^0=1,\; X^1=-i,\; X^2=X^3=0,$$
so that $X^0\bar X^0=X^1\bar X^1 =1$.  The conditions
$$\frac {\partial V}{\partial e_1^a}=0, \qquad \frac {\partial V}{\partial
e_2^a}=0,$$ are fulfilled by $e_1^a=e_2^a=0$.

As a function of $\sigma$ and $\tau$, the potential $V$ is
composed of two pieces $V=A(\sigma, \tau)+B$. $A(\sigma, \tau)$ is
given by  the two first lines in (\ref{pot}), and it  is positive
definite. $B$ is negative and  constant. The point $\sigma=\tau=i$
is an extremum of $V$ and a minimum of $A$.  The value of the
potential at this extremum is
$$V( \sigma=\tau=i)=2\mathrm{e}^{2\phi}\mathrm{e}^{\tilde
K}(g_0^2e_1^ae_1^a+g_1^2e_2^ae_2^a)\geq 0.$$ This implies that the
potential is positive definite for all $\sigma$ and $\tau$.

Summarizing,  we have that $V\geq 0$ and that $V=0$ at the
extrema, so they are minima. The extremum condition does not fix
the scalars $\phi$, $\rho$, $e_3^a$ and the remaining $C^I$ (all
of them  except for $C^m$ with $m=1,2$, which disappear from the
spectrum.)

\bigskip
 The gravitino mass matrix is
\begin{eqnarray*} M_{AB} &=& 2S_{AB}= i  L^\Lambda P_\Lambda^x(\sigma^x)_{AB} =
i
\mathrm{e}^\phi \begin{pmatrix}
L^0 g_0 + i L^1 g_1 &0 \\
0 & -L^0 g_0 + i L^1 g_1 \end{pmatrix}\\
&=& \mathrm{e}^\phi \mathrm{e}^{\frac{\hat K}2} \frac 1{[i(\rho -
\bar \rho )]^{\frac 12}}
\begin{pmatrix}
X^0 g_0 + i X^1 g_1 &0 \\
0 & -X^0 g_0 + i X^1 g_1 \end{pmatrix}
\end{eqnarray*}
At the point $\sigma = \tau = i$ we have $X^0 = 1$ and $X^1 = -i$,
so that the eigenvalues squared are:
$$ m^2_{1,2} = \frac 12 \frac {\mathrm{e}^{2\phi}}{i(\rho -
\bar \rho )} (g_0 \pm g_1)^2,$$ which gives the gravitino masses
measured in terms of the $\K$ volume and the $\T^2$ Kaehler
modulus.

\subsection{ \label{noscale} More general vacua}
More general vacua, preserving $N=1,0$ supersymmetry, can be
obtained by considering an arbitrary vector coupling $g_\Lambda$.
$g_0$ and $g_1$ gauge two of the isometries $C^m$, while $g_2$ and
$g_3$ gauge two of the isometries $C^a$.

Taking the Killing vectors as
$$k^{C^{m=1}}_0 =g_0, \quad k^{C^{m=2}}_1 =g_1, \quad k^{C^{a=1}}_2 =g_2, \quad k^{C^{a=2}}_3
=g_3,$$ we first notice that the condition on the vector multiplet
sector $X^2=X^3=0\Longrightarrow \sigma=\tau=i$ still holds. In
the hypermultiplet sector, the condition for $N=1$ vacua we had,
$e_{m=1,2}^a=0$, is supplemented by the extra condition, coming
from the gauginos variation, $e^{a=1,2}_m=0$. This eliminates from
the spectrum two extra scalars $e^{a=1,2}_3=0$ together with the
axions $C_{a=1,2}$. We are therefore left 18 chiral multiplets
from the hypermultiplet sector, no massless vector multiplets and
one chiral multiplet from the $N=2$ vector multiplet sector.

If we relax the condition $|g_0|=|g_1|$, the vacua will not
preserve any supersymmetry, but still will have vanishing vacuum
energy, as can be shown by looking at the scalar potential.

We want to note the close connection of the present $N=2$ model
with another $N=2$ model more recently discussed as an effective
theory for $N=2$ vacua of the $\T^6/\Z_2$ orientifold
\cite{adflq}. The vector multiplet sector of that theory is
obtained by ``higgsing" two of the three vector multiplets without
breaking $N=2$ supersymmetry, as discussed in Section \ref{n=2}.
The scalars in the remaining hypermultiplets parametrize the
manifold $\rSO(4,18)/(\rSO(4)\times \rSO(18))$, but since the
vector multiplet sector is the the same as in the $T^6/\Z_2$
truncated model of Ref. \cite{adflq}, the pattern of the
supersymmetry breaking is very similar, and is insensitive to the
number the of hypermultiplets. This is because the relation
$$P_\Lambda^xP_\Sigma^x(U^{\Lambda\Sigma}-3\bar{L}^\Lambda
L^\Sigma)=-2P_\Lambda^xP_\Sigma^x\bar{L}^\Lambda L^\Sigma$$ is
still satisfied as in the model of Ref. \cite{adflq}.

 This model
was also shown to be connected to the minimal $N=2$ model studied
in Refs. \cite{cgp,fgp}. The vanishing potential of the theory in
Ref. \cite{adflq} was closely connected to the positive potential
of the theory in Refs. \cite{cgp,fgp}, which lead to moduli
stabilization, as expected from the $\T^6/\Z_2$ orientifold
analysis \cite{gkp,fp,kst,lo,fm,kstt}.

\subsection{ N=1 $\to$ N=0 no scale supergravities}
In this section we find, by truncation from $N=2$, a $N=1$ theory
with a transition $N=1\to N=0$ \cite{fpo}. The truncation $N=2\to
N=1$ can be formally obtained by integrating out the second
gravitino multiplet, together with the states which receive mass
in the $N=2 \to N=1$ phase transition. Twenty chiral multiplets
from the hypermultiplet sector remain massless in the truncation
together with one chiral multiplet from the vector multiplet
sector.

On the other hand, by relaxing the condition $|g_0|=|g_1|$, which
makes the transition $N=1 \to N=0$, none of the scalars in the
$N=1$ theory take mass. This means that the $N=1 \to N=0$
transition must occur with a vanishing potential. To understand
this, we analyze the moduli space of the 21 chiral multiplets. The
left over moduli from the quaternionic manifold parametrize the
submanifold
$$\frac {\rSO(2,20)}{\rSO(2)\times \rSO(20)},$$
which is a Kaehler-Hodge manifold with Kaehler potential
\begin{equation}K_1=-\ln[(x_0+\bar x_0)^2-\sum_{a=1}^{19}(x_a+\bar
x_a)^2],\label{k1}\end{equation} where
$$x^a=\frac 12(e^a_3+iC^a), \qquad x_0=\frac 12(\mathrm{e}^{-\phi}+iC^{m=3}),$$
and $(\phi, e^a_3)$ parametrize the moduli space of the metrics
$$\frac {\rSO(1,19)}{\rSO(19)}\times \R^+_{\T^2}.$$
The remaining chiral multiplet contains the $\K$ volume modulus
with Kaehler potential $\tilde K= -\ln i(\rho-\bar \rho)$. The
total Kaehler potential of the manifold of the scalars in the
chiral multiplets is a cubic polynomial
$$K=-\ln i(\rho-\bar \rho)[(x_0+\bar x_0)^2-\sum_{a=1}^{19}(x_a+\bar
x_a)^2].$$ The flux which breaks $N=1$ to $N=0$ corresponds to a
constant superpotential $W=a=\mathrm{constant}$. In this situation
the scalar potential \cite{cfgp},
$$V=\mathrm{e}^K(\mathcal{D}_iW\mathcal{D}_{\bar j}\bar W G^{i\bar
j}-3|W|^2),\qquad  G^{i\bar j}=(\partial_i\partial_{\bar
j}K)^{-1},\quad \mathcal{D}_iW=(\partial_i+\partial_iK)W,$$ is
identically zero \cite{cfkn,ckpdfwg,bcf}. The gravitino mass is
related to the overall volume of $\K\times \T^2$,
$$m^2_{3/2}=\mathrm{e}^K a^2.$$
This is a standard $N=1$ no scale model. However, it is different
from the one obtained by $\T^6/\Z_2$ compactification because it
has a much richer structure of moduli. The goldstino is
essentially the fermion superpartner of the overall $\K \times
\T^2$ volume while all the other fermions receive a mass due to
the flux, equal to the gravitino mass \cite{cfkn}.

Note that if instead of twenty  chiral multiplets we had
considered a model with two chiral multiplets in (\ref{k1}), we
would have retrieved the analysis of \cite{fpo}.

\section{Concluding remarks}

In the present investigation we have shown that compactifications
on $\K \times \T^2 / \Z_2$ orientifold can be reproduced by a
gauged $N=2$ supergravity which exactly gives the same $N=2,1,0$
vacua as obtained by analyzing the existence of the ten
dimensional supergravity solution.

The choice of the gauging is the crucial issue. The existence of
backgrounds with vanishing vacuum energy and broken supergravity
closely depends on the fact that the corresponding gauging
requires a choice of symplectic sections for special geometry
which do not admit a prepotential. This is required in order to
evade a no-go theorem on partial breaking of supersymmetry
\cite{mp}.

This analysis can be generalized by including Yang-Mills degrees
of freedom coming from the branes, as well as more general fluxes
related to the supergravity charges $f_{m\Lambda}$,
$h_{a\Lambda}$.

Our analysis extends previous studies on partial super-Higgs in
$N=2$ supergravity considered in the literature
\cite{ckpdfwg,fkpz,fklz,fgpt}. In particular the no-scale
structure is closely related to the minimal model \cite{cgp,fgp}
and it only depends on universal properties leading to
cancellation of positive and negative contributions in the scalar
potential as it occurred in $N=1$ no-scale models
\cite{cfkn,elnt}.

Another interesting problem which is left aside here, is the
effect of the quantum corrections in these no-scale models. Some
work along these directions has recently appeared in the
literature \cite{bbhl,kklt}.

\section*{Acknowledgments}
We acknowledge interesting discussions with C. Angelantonj, I.
Antoniadis, C. Kounnas and T. Taylor.

 Work supported in part by  the European
Community's Human Potential Program under contract
HPRN-CT-2000-00131 Quantum Space-Time, in which L. A.,  R. D. and
M. A. Ll. are associated to Torino University.

The work of S. F. has also  been supported by the D.O.E. grant
DE-FG03-91ER40662, Task C.

The work of M. A Ll. has also been supported by the research grant
BFM 2002-03681 from the Ministerio de Ciencia y Tecnolog\'{\i}a
(Spain) and from EU FEDER funds.

\end{document}